\documentclass[aps,prb,twocolumn,superscriptaddress,floatfix]{revtex4}
\usepackage{graphicx,graphics}
\usepackage{dcolumn}
\usepackage{amsmath,amssymb,amsfonts}
\usepackage{latexsym,verbatim}
\usepackage{bm}
\usepackage{color}
\usepackage{ulem}
\usepackage[breaklinks=true,colorlinks,citecolor=blue,linkcolor=blue,urlcolor=blue]{hyperref}
\begin{document}
\title{Electron-hole pairing in graphene-GaAs heterostructures}
\author{A. Gamucci}
\affiliation{NEST, Istituto Nanoscienze-CNR and Scuola Normale Superiore, I-56126 Pisa, Italy}
\author{D. Spirito}
\affiliation{NEST, Istituto Nanoscienze-CNR and Scuola Normale Superiore, I-56126 Pisa, Italy}
\author{M. Carrega}
\affiliation{NEST, Istituto Nanoscienze-CNR and Scuola Normale Superiore, I-56126 Pisa, Italy}
\author{B. Karmakar}
\affiliation{NEST, Istituto Nanoscienze-CNR and Scuola Normale Superiore, I-56126 Pisa, Italy}
\author{A. Lombardo}
\affiliation{Engineering Department, University of Cambridge, Cambridge, CB3 OFA, UK}
\author{M. Bruna}
\affiliation{Engineering Department, University of Cambridge, Cambridge, CB3 OFA, UK}
\author{A.C. Ferrari}
\affiliation{Engineering Department, University of Cambridge, Cambridge, CB3 OFA, UK}
\author{L.N. Pfeiffer}
\affiliation{Department of Electrical Engineering, Princeton University, Princeton, NJ, USA}
\author{K.W. West}
\affiliation{Department of Electrical Engineering, Princeton University, Princeton, NJ, USA}
\author{M. Polini}
\email{m.polini@sns.it}
\affiliation{NEST, Istituto Nanoscienze-CNR and Scuola Normale Superiore, I-56126 Pisa, Italy}
\author{V. Pellegrini}
\email{Vittorio.Pellegrini@iit.it}
\affiliation{Istituto Italiano di Tecnologia, Graphene labs, Via Morego 30, I-16163 Genova, Italy}
\affiliation{NEST, Istituto Nanoscienze-CNR and Scuola Normale Superiore, I-56126 Pisa, Italy}
\begin{abstract}
Vertical heterostructures combining different layered materials offer novel opportunities for applications~\cite{novoselov_ps_2012,bonaccorso_matertoday_2012,britnell_science_2012} and fundamental studies of collective behavior driven by inter-layer Coulomb coupling~\cite{ponomarenko_naturephys_2011,kim_prb_2011,gorbachev_naturephys_2012,kim_ssc_2012}. Here we report heterostructures comprising a single-layer (or bilayer) graphene carrying a fluid of massless (massive) chiral carriers~\cite{andre_naturemater_2007}, and a quantum well created in GaAs 31.5~nm below the surface, supporting a high-mobility two-dimensional electron gas. These are a new class of double-layer devices composed of spatially-separated electron and hole fluids. We find that the Coulomb drag resistivity significantly increases for temperatures below 5-10~K, following a logarithmic law. This anomalous behavior is a signature of the onset of strong inter-layer correlations, compatible with the formation of a condensate of permanent excitons. The ability to induce strongly-correlated electron-hole states paves the way for the realization of coherent circuits with minimal dissipation~\cite{high_science_2008,kuznetsova_opticsletters_2010,sanvitto_naturecommun_2013} and nanodevices including analog-to-digital converters~\cite{dolcini_prl_2010} and topologically protected quantum bits~\cite{peotta_prb_2011}.
\end{abstract}

\maketitle

\begin{figure}[t]
\center{
\includegraphics[width=\columnwidth]{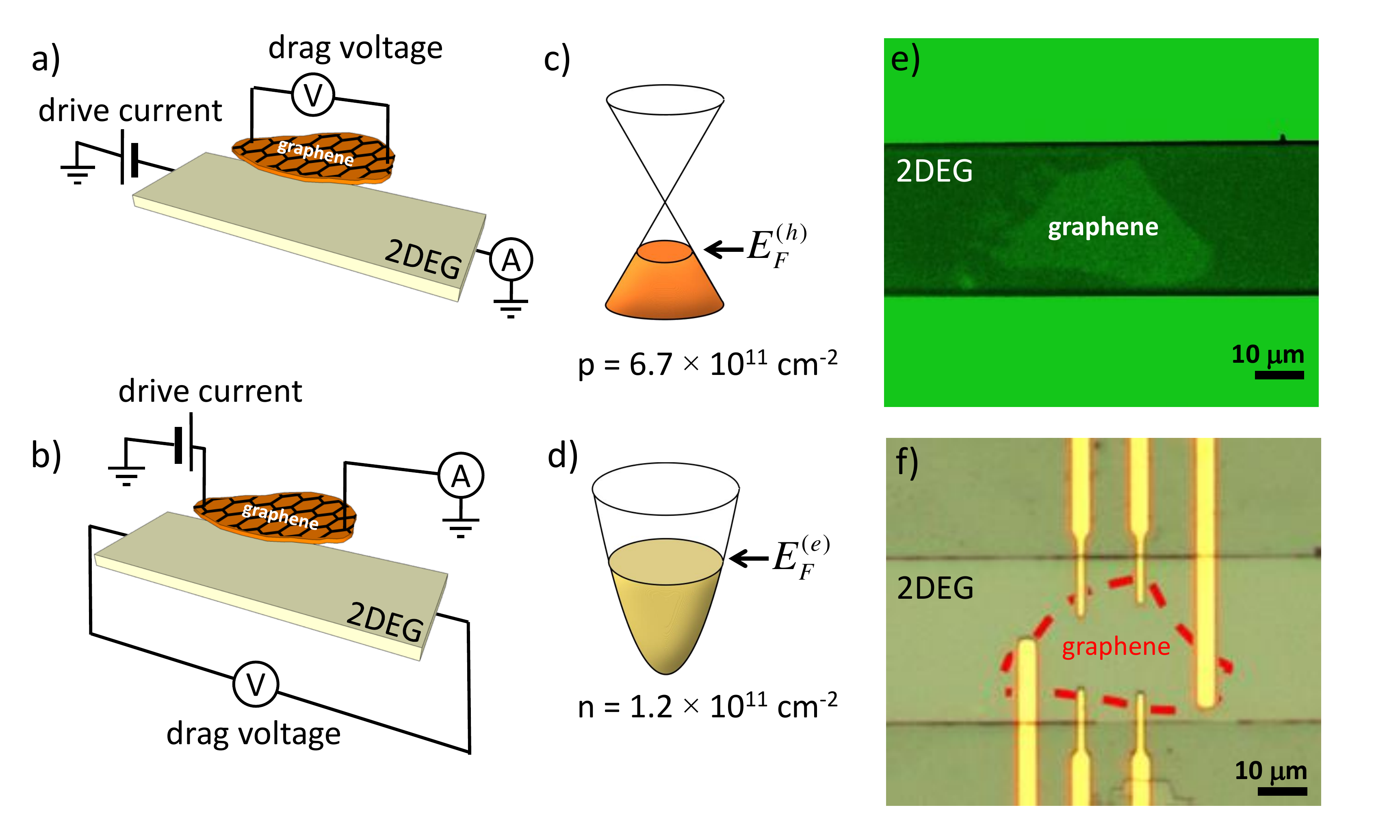}
\caption{{\bf Hybrid SLG/2DEG devices and Coulomb drag transport setup.} a,b) Configurations for Coulomb drag measurements. In a), a voltage drop $V_{\rm drag}$ appears in graphene, in response to a drive current $I_{\rm drive}$ flowing in the 2DEG. In b) the opposite occurs. The drag voltage is measured with a low-noise voltage amplifier coupled to a voltmeter as a function of the applied bias. The drive current is also monitored. c) Conical massless Dirac fermion band structure of low-energy carriers in SLG~\cite{andre_naturemater_2007}. The SLG in this work is hole doped. d) Parabolic band structure of ordinary Schr\"{o}dinger electrons in the 2DEG. e) Optical micrograph of the device prior to the deposition of Ohmic contacts. The SLG flake becomes visible in green light after the sample is coated with a polymer (PMMA)~\cite{cnano}. f) Optical microscopy image of the contacted SLG  on the etched 2DEG GaAs channel. The red dashed line denotes the SLG boundaries.\label{fig:one}}
}
\end{figure}

Our vertical heterostructures are prepared as follows. Single-layer (SLG) and bilayer graphene (BLG) flakes are produced by micromechanical exfoliation of graphite on Si/${\rm SiO}_2$~\cite{kostya_pnas_2005}. The number of layers is identified by a combination of optical microscopy~\cite{cnano} and Raman spectroscopy~\cite{RamanACF,RamanACF2}. The latter is also used to monitor the sample quality by measuring the D to G ratio~\cite{cancado} and the doping level~\cite{das}. Selected flakes are then placed onto a GaAs-based substrate at the center of a pre-patterned Hall bar by using a polymer-based wet transfer process~\cite{bonaccorso_matertoday_2012} (see Appendix~\ref{appendix:sample} for further details). The GaAs-based substrates consist of modulation-doped GaAs/AlGaAs heterostructures hosting a two-dimensional electron gas (2DEG) in the GaAs quantum well placed $31.5~{\rm nm}$ below the surface. The heterostructures are grown by molecular beam epitaxy~\cite{loren}, and consist of a $n$-doped GaAs cap layer, a AlGaAs barrier, a GaAs well and a thick AlGaAs barrier with a delta doping layer (see Appendix~\ref{appendix:sample} for further details). Two different samples are fabricated: sample A having a $15~{\rm nm}$-thick quantum well and sample B with a $22~{\rm nm}$-thick quantum well. Hall bars ($300~\mu{\rm m}$ wide and $1500~\mu{\rm m}$ long) are fabricated by UV lithography. Ni/AuGe/Ni/Au layers are then evaporated and annealed at $400^\circ~{\rm C}$ to form Ohmic contacts to the 2DEG, to be used for transport and Coulomb drag measurements (see Fig.~\ref{fig:one}). The Hall bar mesas are defined by conventional wet etching in acid solution. To ensure that the current in the 2DEG flows only in the region below the graphene flakes, channels with a width comparable to the transferred graphene flakes (typically $\sim 30~\mu{\rm m}$), are defined in the Hall bar by means of electron beam lithography and wet etching, Figs.~\ref{fig:one}e)-f). A SLG flake is transferred onto sample A and a BLG flake onto sample B. The integrity of the SLG and BLG flakes is monitored throughout the process by Raman spectroscopy. Fig.~\ref{fig:Stwo} in Appendix~\ref{appendix:sample} compares the Raman spectra of as prepared SLG on Si/SiO$_2$ and after transfer on the GaAs substrate. Analysis of G peak position, Pos(G), its full width at half maximum, FWHM(G), Pos(2D) and the area and intensity ratios of 2D and G peaks, allow us to monitor the amount and type of doping~\cite{RamanACF2,das,basko}. This indicates a small $p$ doping for the as-prepared sample, decreasing to below $100~{\rm meV}$ for the transferred sample~\cite{RamanACF2,das,basko}. The absence of a significant D peak both before and after transfer indicates that the samples have negligible amount of defects~\cite{cancado,RamanACF2} and that the transfer procedure does not add defects. Similarly, no increase in defects is seen for the BLG samples.

To ensure that the two-dimensional (2d) hole gas in SLG/BLG and the 2DEG in GaAs are electrically isolated, we monitor the inter-layer $I_{\rm I}$-$V_{\rm I}$ characteristics in the $0.25~{\rm K}$-$50~{\rm K}$ temperature range (see Appendix~\ref{appendix:interlayertransport}), with $I_{\rm I}$ and $V_{\rm I}$ the inter-layer (``leakage'') current and inter-layer voltage, respectively, and the layers being the SLG (or BLG) and the GaAs quantum well. In SLG-based devices, a negligible inter-layer current $<0.2~{\rm nA}$ is measured for $V_{\rm I}$ up to $-0.3~{\rm V}$ for all temperatures, leading to inter-layer resistances~$\sim 1~{\rm G}\Omega$. In the case of BLG, for $T \sim 45~{\rm K}$, $I_{\rm I}$ increases to $100~{\rm nA}$ at $V_{\rm I}=-0.3~{\rm V}$, with the inter-layer resistance increasing to several ${\rm M}\Omega$. In all cases, therefore, the inter-layer resistance is much larger than the {\it largest} intra-layer resistance for SLG, BLG and GaAs quantum well, which is $\sim 10~{\rm k}\Omega$.

To search for signatures of correlations between the 2DEG in the GaAs quantum well and the chiral hole fluid~\cite{andre_naturemater_2007} in SLG or BLG, we measure the temperature dependence of the {\it Coulomb drag} resistance $R_{\rm D}$. In a Coulomb drag experiment~\cite{gramila_prl_1991,sivan_prl_1992,rojo_jpcm_1999} a current source is connected to one of the two layers (the {\it active} or {\it drive} layer). The other layer (the {\it passive} layer) is connected to an external voltmeter, so that the layer can be assumed to be an open circuit (no current can flow in it). The drive current $I_{\rm drive}$ drags carriers in the passive layer, which accumulate at the ends of the layer, building up an electric field. The voltage drop $V_{\rm drag}$ related to this field is then measured. The quantity $R_{\rm D}$ is defined as the ratio $V_{\rm drag}/I_{\rm drive}$ and is determined by the rate at which momentum is transferred between quasiparticles in the two layers~\cite{rojo_jpcm_1999}.

Since $R_{\rm D}$ originates from electron-electron interactions, it contains information on many-body effects stemming from correlations~\cite{zheng_prb_1993,kamenev_prb_1995}. Experimentally, Coulomb drag has been indeed used as a sensitive probe of transitions to the superconducting state~\cite{giordano_prb_1994}, metal-insulator transitions~\cite{pillarisetty_prb_2005}, transition to the Wigner crystal phase in quantum wires~\cite{yamamoto_science_2006}, and exciton condensation in quantum Hall bilayers~\cite{nandi_nature_2012}. In the context of spatially-separated systems of electrons and holes, in the absence of a magnetic field, the theoretical studies in Ref.~\onlinecite{vignale_prl_1996} indicated that $R_{\rm D}$ is comparable to the isolated layer resistivity in the exciton condensed phase of an electron-hole (e-h) double layer. For $T$ larger than, but close to, the mean-field critical temperature $T_{\rm c}$, the occurrence of e-h pairing fluctuations increases $R_{\rm D}$ with respect to its value in the Fermi-liquid phase~\cite{hu_prl_2000,snoke_science_2002,mink_prl_2012,mink_prb_2013}. An increase in $R_{\rm D}$ with a suitable functional dependence on $T$ indicates the transition to an exciton condensate~\cite{hu_prl_2000,snoke_science_2002,mink_prl_2012,mink_prb_2013}. This is similar to the enhancement of conductivity in superconductors due to Cooper-pair fluctuations (``paraconductivity'') above, but close to, the critical temperature~\cite{Larkin_and_Varlamov}.

\begin{figure}[t]
\center{
\includegraphics[width=\columnwidth]{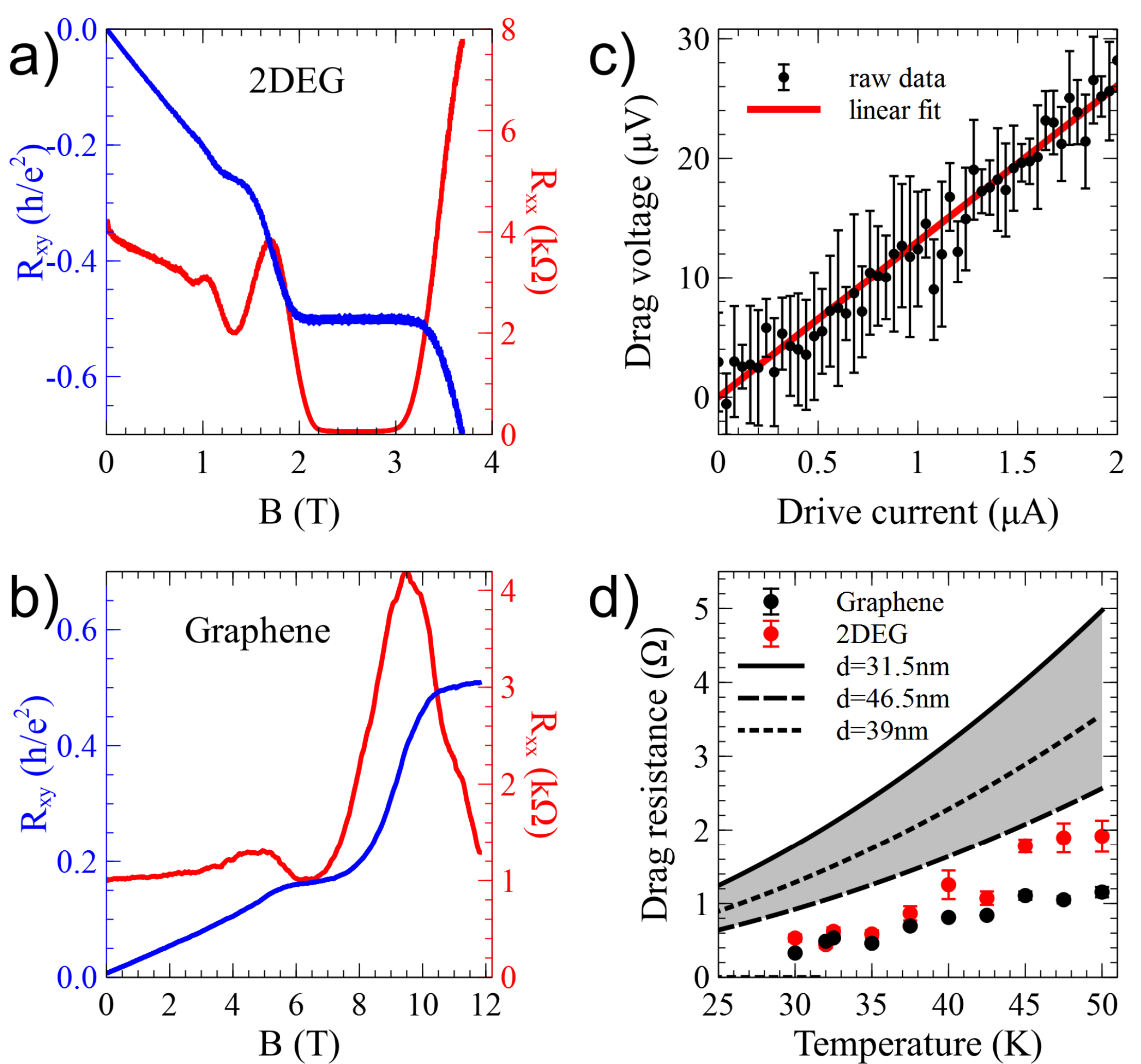}
\caption{{\bf Magneto-transport characterization of the 2DEG and SLG and high-temperature Coulomb drag.} a,b) Hall (blue solid line) and longitudinal (red solid line) resistance of 2DEG and SLG, respectively. Hall measurements are performed in the two layers with the same configuration of electrical connections: Hall resistance is positive for holes and negative for electrons. c) Drag voltage in the 2DEG as a function of the drive current flowing in SLG at $T=42.5~{\rm K}$: data and a linear fit are shown. Error bars are calculated as standard deviations from the average 
of $10$ current sweeps. d) Drag resistance as a function of temperature. Black (red) points refer to $R_{\rm D}$ derived by measuring the voltage drop in SLG (2DEG), respectively. The three lines are Boltzmann-transport calculations in the Fermi-liquid regime (see Ref.~\onlinecite{principi_prb_2012} and Appendix~\ref{appendix:theory}). Different curves refer to different values of the inter-layer distance $d$: $d = 31.5~{\rm nm}$ (solid line), $46.5~{\rm nm}$ (long-dashed line), and $39~{\rm nm}$ (short-dashed line).\label{fig:two}}
}
\end{figure}

Prior to Coulomb drag experiments, we perform magneto-transport measurements at $4~{\rm K}$, as for Figs.~\ref{fig:two}a)-b). In our setup, the 2DEG is induced in the quantum well by shining light from an infrared diode. In the SLG/2DEG device we find a 2DEG with density $n =1.2 \times 10^{11}~{\rm cm}^{-2}$ from low-field (below 1 Tesla) classical Hall effect and a mobility $\mu_{\rm e}=13000~{\rm cm}^{2}/({\rm V} {\rm s})$ at 4K. At $T=45~{\rm K}$, the density decreases to $4.0\times 10^{10}~{\rm cm}^{-2}$ and $\mu_{\rm e} = 8700~{\rm cm}^2/({\rm V} {\rm s})$. Fig.~\ref{fig:two}a) shows the quantum Hall effect in the 2DEG. The quantum Hall plateaus at $h/(2 e^2)$ and $h/(4 e^2)$ (blue trace), correspond to the first two spin degenerate Landau levels~\cite{prange}. In correspondence of the plateaus, minima are found~\cite{prange} in the longitudinal resistance $R_{xx}$ (red trace).

The 2d hole fluids in SLG and BLG have their highest mobility when the 2DEG is {\it not} induced. This is shown in Fig.~\ref{fig:two}b) for the SLG-based device (see also Appendix~\ref{appendix:characterization}). Figs.~\ref{fig:two}a)-b) indicate that the sign of the Hall resistance $R_{xy}$ in SLG is opposite to the 2DEG, thereby demonstrating that SLG is $p$-doped. At $4~{\rm K}$ the hole density is $p=9.9 \times 10^{11}~{\rm cm}^{-2}$ and $\mu_{\rm h} = 4100~{\rm cm}^2/({\rm V} {\rm s})$. At $45~{\rm K}$ the corresponding values are $p=6.7 \times 10^{11}~{\rm cm}^{-2}$ and $\mu_{\rm h} = 2400~{\rm cm}^2/({\rm V} {\rm s})$. Low-temperature magneto-transport in SLG, Fig.~\ref{fig:two}b), reveals quantum Hall plateaus at $h/(2e^2)$ and $h/(6 e^2)$, corresponding to massless Dirac fermions with spin and valley degeneracy~\cite{andre_naturemater_2007}. On the contrary, when the 2DEG is optically induced, the hole density in SLG at $4~{\rm K}$ is $p = 6.7 \times 10^{11}~{\rm cm}^{-2}$ and $\mu_{\rm h} = 2100~{\rm cm}^2/({\rm V} {\rm s})$, thereby weakening the manifestations of the quantum Hall effect (see Appendix~\ref{appendix:characterization}). The degradation of the SLG transport properties in the presence of the 2DEG could be linked to the creation of ionized Si donors within the $n$-doped GaAs cap layer, acting as positively-charged scatterers~\cite{dassarma_rmp_2011}.

We now focus on the Coulomb drag experiments. These are performed in the configuration sketched in Figs.~\ref{fig:one}a)-b) and in a $^3$He cryostat with a $240~{\rm mK}$-$50~{\rm K}$ temperature range. Ten $V_{\rm drag}$ - $I_{\rm drive}$ curves in a dc configuration are acquired for each $T$ and then averaged. We first address the SLG/2DEG case. Fig.~\ref{fig:two}c) reports a representative set of averaged drag voltage data taken in the 2DEG at $T = 42.5~{\rm K}$. In this configuration, the SLG gating effect and consequent carrier depletion in the 2DEG are avoided by applying a positive current, from 0 to $+2$~$\mu$A in the SLG channel. Fig.~\ref{fig:two}c) shows that at this representative value of $T$ the drag voltage is linear with the drive current, thereby allowing the extraction of $R_{\rm D}$ from the slope of a linear fit.

\begin{figure}[t]
\center{
\includegraphics[width=\columnwidth]{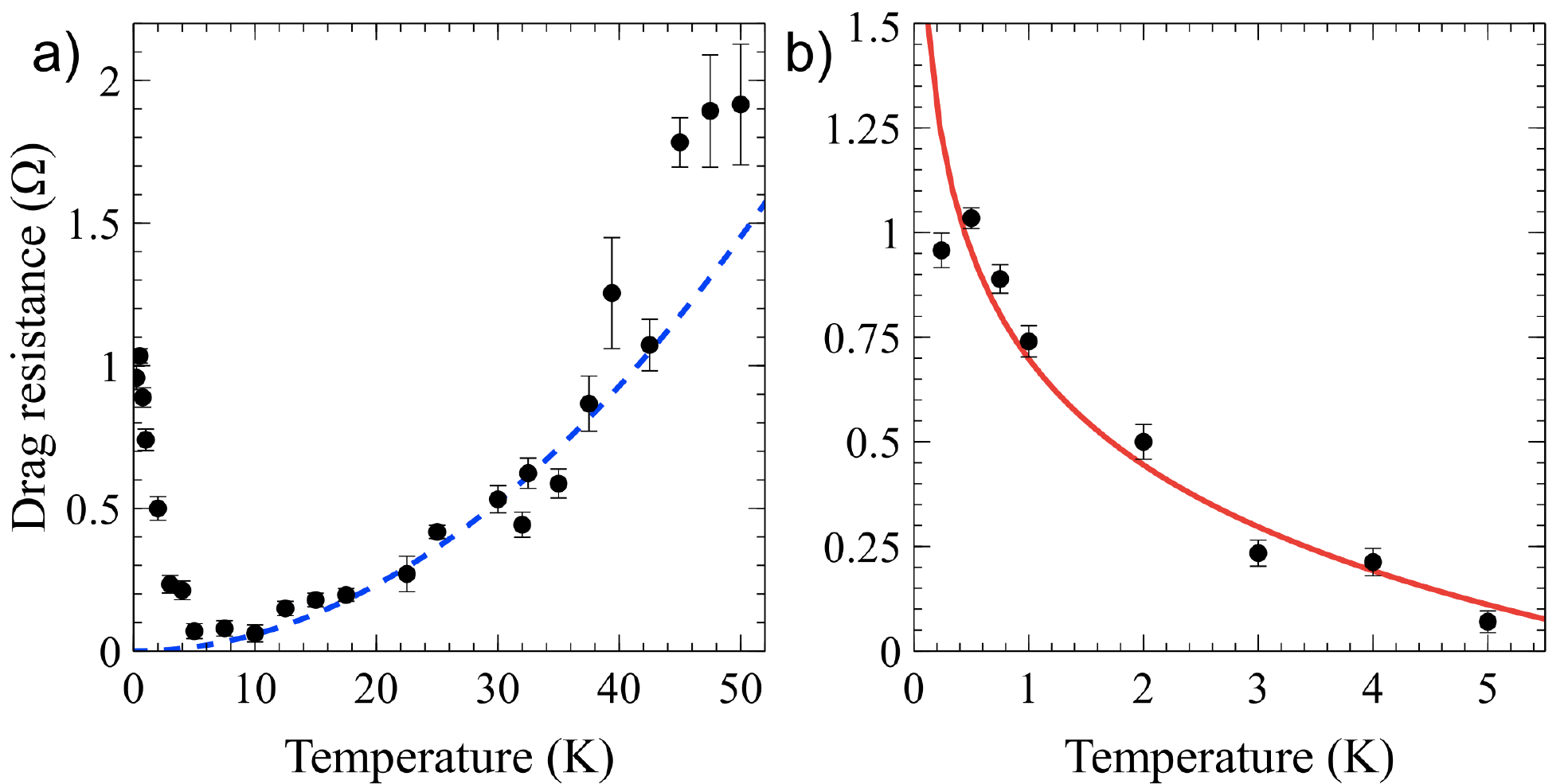}
\caption{{\bf Temperature dependence of the Coulomb drag resistance in the SLG/2DEG vertical heterostructure.} a) $R_{\rm D}$ obtained by measuring the voltage drop in the 2DEG (passive layer) in response to a drive current flowing in the SLG (drive layer). The dashed blue line is a best-fit of the standard Fermi-liquid type~\cite{rojo_jpcm_1999}: $R_{\rm D}(T) = a T^2$ with $a=  (5.8 \pm 0.3) \times 10^{-4}~\Omega/{\rm K}^2$. b) Zoom of $R_{\rm D}$ in the low-$T$ limit. The red solid line is a fit based on the functional form reported in Eq.~(\ref{eq:logfit}). This fit describes very well the $R_{\rm D}$ upturn at low $T$ as the system approaches $T_{\rm c} \sim 10~{\rm mK}$-$100~{\rm mK}$.\label{fig:three}}
}
\end{figure}

Fig.~\ref{fig:two}d) plots $R_{\rm D}$ for $30~{\rm K}\leq T\leq 50~{\rm K}$, with the 2DEG used as the drive (black points) or passive (red points) layer. It also reports calculations of the $T$ dependence of $R_{\rm D}$ in a hybrid Dirac/Schr\"{o}dinger SLG/2DEG double layer within a Boltzmann-transport theory, which is justified in the Fermi-liquid regime~\cite{carrega_njp_2012,principi_prb_2012}. This is done by generalizing the theory of Ref.~\onlinecite{principi_prb_2012} to include effects due to the finite width of the GaAs quantum well (see Appendix~\ref{appendix:theory}). This shows that the experimental results in this temperature range are consistent with the canonical Fermi-liquid prediction~\cite{gramila_prl_1991,sivan_prl_1992,rojo_jpcm_1999,zheng_prb_1993,kamenev_prb_1995,carrega_njp_2012,principi_prb_2012}, i.e.~$R_{\rm D} \propto T^2$---see also Fig.~\ref{fig:three}a)---as constrained by the available phase-space of the initial and final states involved in the scattering process. The magnitude of the measured effect, however, is smaller than predicted by theory. Discrepancies of similar magnitude have been previously reported for Coulomb drag measurements between two SLG encapsulated in hexagonal Boron Nitride~\cite{gorbachev_naturephys_2012}. Fig.~\ref{fig:two}d) demonstrates that the Onsager reciprocity relations~\cite{onsager_physrev_1931_I,onsager_physrev_1931_II}, which in our case require that the resistance measured by interchanging drive and passive layers should be unchanged, are satisfied in the $30~{\rm K}\leq T\leq 40~{\rm K}$ range. A slight violation of reciprocity, whose origin is at present not understood, seems to occur for $T>40~{\rm K}$.

We now discuss the behavior of $R_{\rm D}$ in the low-$T$ regime. We follow Ref.~\onlinecite{gorbachev_naturephys_2012} and use the lowest quality layer, in our case SLG, as the drive layer and measure the drag voltage in the 2DEG. In the reversed configuration, the drag voltage measured in SLG displays fluctuations~\cite{gorbachev_naturephys_2012,kim_ssc_2012} as a function of the drive current, which hamper the extraction of $R_{\rm D}$, see Appendix~\ref{appendix:fluctuations}. $R_{\rm D}$ measured in the 2DEG reveals an anomalous behavior below $10~{\rm K}$. Fig.~\ref{fig:three} indicates that $R_{\rm D}$ deviates from the ordinary $T^2$ dependence, as shown by a large upturn for $T$ lower than an ``upturn'' temperature $T_{\rm u} \sim 5~{\rm K}$. The enhancement of $R_{\rm D}$ at low $T$ is a very strong effect: the drag signal increases by more than one order of magnitude by decreasing $T$ below $T_{\rm u}$, where $R_{\rm D}$ is vanishingly small, in agreement with Fermi-liquid predictions (see Appendix~\ref{appendix:theory}), down to $T = 240~{\rm mK}$.

Fig.~\ref{fig:three}b) is a zoom of the drag enhancement data in the low-$T$ range together with a fit (solid line) of the type~\cite{mink_prl_2012,mink_prb_2013}:
\begin{equation}\label{eq:logfit}
R_{\rm D}(T) = R_0 + A \log\left(\frac{T_{\rm c}}{T - T_{\rm c}}\right)~,
\end{equation}
where $R_0$ and $A$ are two fitting parameters and $T_{\rm c}$ is the mean-field critical temperature of a low-$T$ phase transition. Even though this fitting procedure cannot predict $T_{\rm c}$, it is in excellent agreement~\cite{footnote_temperature} with the data for $T_{\rm c}$ in the range $10~{\rm mK}$-$100~{\rm mK}$.

The logarithmic enhancement of $R_{\rm D}$ described by Eq.~(\ref{eq:logfit}) was theoretically predicted in Refs.~\onlinecite{mink_prl_2012,mink_prb_2013} on the basis of a Boltzmann transport theory for e-h double layers, where the scattering amplitude is evaluated in a ladder approximation~\cite{Larkin_and_Varlamov}. Similar results were obtained on the basis of a Kubo-formula approach~\cite{hu_prl_2000}. Within these theoretical frameworks, the enhancement is attributed to e-h pairing fluctuations~\cite{hu_prl_2000,snoke_science_2002,mink_prl_2012,mink_prb_2013,rist_prb_2013} extending above $T_{\rm c}$ for a phase transition into an exciton condensed phase. This is ascribed to the quasi-2d nature of our SLG/2DEG heterostructure and shares similarities with other 2d systems where fluctuations play an important role like cuprate superconductors (see, for example, Ref.~\onlinecite{zhang_prb_2013}) and cold Fermi gases~\cite{feld_nature_2011}.

\begin{figure}[t]
\center{
\includegraphics[width=\columnwidth]{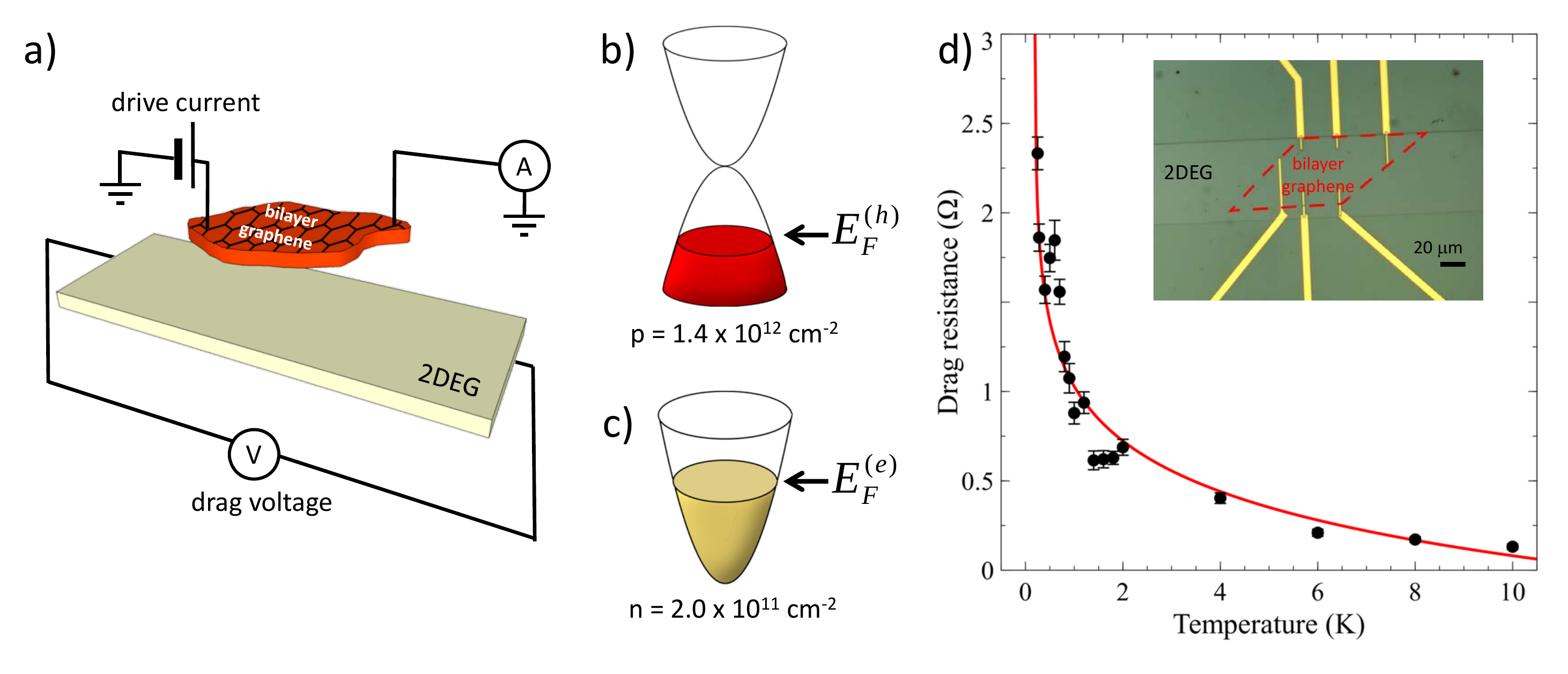}
\caption{{\bf Temperature dependence of $R_{\rm D}$ in the BLG/2DEG vertical heterostructure.} a) Configuration for the Coulomb drag measurements. A voltage drop $V_{\rm drag}$ appears in the 2DEG in response to a current $I_{\rm drive}$ that flows in BLG.  b) Low-energy parabolic  band structure of massive chiral holes in BLG~\cite{andre_naturemater_2007}. c) Parabolic band structure of Schr\"{o}dinger electrons in the 2DEG. d) $R_{\rm D}$ in the low-$T$ limit. The red solid line is a fit based on the functional form reported in Eq.~(\ref{eq:logfit}) with $T_{\rm c} \sim 190~{\rm mK}$. The inset shows an optical microscopy image of the contacted BLG on the etched 2DEG GaAs channel. The red dashed line denotes the BLG boundaries.\label{fig:four}}
}
\end{figure}

To further investigate this effect we explore a second device comprising of a hole-doped exfoliated BLG deposited on the surface of a GaAs quantum heterostructure. The hole density in BLG is $p =1.4 \times 10^{12}~{\rm cm}^{-2}$ from the low-field (below $1~{\rm Tesla}$) classical Hall effect and the mobility is $670~{\rm cm}^{2}/({\rm V} {\rm s})$ at $4~{\rm K}$. The 2DEG has an electron density $n =2 \times 10^{11}~{\rm cm}^{-2}$ and a mobility $86000~{\rm cm}^{2}/({\rm V} {\rm s})$ at $4~{\rm K}$. Contrary to the SLG/2DEG case, in the BLG/2DEG device both electron and hole fluids have parabolic energy-momentum dispersions, Figs.~\ref{fig:four}b)-c), recently predicted~\cite{perali_prl_2013} to be particularly favorable for the occurrence of e-h pairing. Since e-h pairing stems from e-e interactions, a lower kinetic energy in BLG (vanishing like $k^2$~\cite{andre_naturemater_2007} rather than like $k$ for small values of momentum $\hbar k$) compared to SLG enhances the {\it relative} importance of Coulomb interactions~\cite{Giuliani_and_Vignale} in BLG/2DEG heterostructures. To probe this, we measure the evolution of $R_{\rm D}$ as a function of $T$ using BLG as the drive layer. Fig.~\ref{fig:four} again shows a significant departure from the Fermi-liquid $T^2$ dependence. Consistent with the expected larger impact of interactions~\cite{perali_prl_2013,Giuliani_and_Vignale}, we get $T_{\rm u} \sim 10~{\rm K}$, i.e.~twice the SLG/2DEG case, while the best fit of $R_{\rm D}$ data based on Eq.~(\ref{eq:logfit}) yields $T_{\rm c} = 190~{\rm mK}$ (to be compared with $T_{\rm c} = 10~{\rm mK}$-$100~{\rm mK}$ in the SLG/2DEG).

A possible approach to further increase $T_{\rm u}$ and $T_{\rm c}$ is to tune the electron $n$ and hole $p$ densities in the two layers in such a way to match the corresponding Fermi wave numbers, $k^{({\rm e})}_{\rm F}$ and $k^{({\rm h})}_{\rm F}$, respectively. In our devices the mismatch in the Fermi wave numbers is $\Delta k_{\rm F} \equiv (k^{({\rm h})}_{\rm F} - k^{({\rm e})}_{\rm F})/(k^{({\rm h})}_{\rm F} + k^{({\rm e})}_{\rm F})\sim 25\%$ in the SLG/2DEG and $\Delta k_{\rm F} \sim 30\%$ in the BLG/2DEG case. Such a mismatch is expected~\cite{pieri_prb_2007} to weaken the robustness of the exciton condensate phase in which condensed e-h pairs have zero total momentum $\hbar K$. Preliminary calculations~\cite{perali_private_commun}, which include screening in the condensed phase~\cite{sodemann_prb_2012,lozovik_prb_2012,neilson_arxiv_2013}, indicate that the $K =0$ exciton condensate state persists even in the presence of these values of $\Delta k_{\rm F}$, with $T_{\rm c}$ scales comparable to those reported here. On the other hand, a mismatch in the Fermi wave numbers of the two fluids may favor other superfluid states, such as those discussed in Refs.~\onlinecite{fulde_pr_1964,larkin_jept_1965,casalbuoni_rmp_2004,pieri_prb_2007,subasi_prb_2010}. These states are however rather fragile in dimensionality $d > 1$, although some evidence was reported, e.g., in the layered heavy-fermion superconductor ${\rm  CeCoIn}_5$~\cite{bianchi_prl_2003}.

The topic of exciton condensation is at the front-end of current condensed-matter research and it involves experimental studies of a wide class of solid-state systems~\cite{rontani}. Those include exciton-polaritons in semiconductor microcavities~\cite{Kasprzak}, which however display ultrashort (picosecond) lifetimes and optically-created indirect excitons in asymmetric semiconductor double quantum wells, where condensation competes with diffusion of the photo-created electrons and holes~\cite{butov}. Condensation of permanent inter-layer excitons was instead demonstrated in electron-electron double layers but at the price of applying high (several Tesla) magnetic fields to enter the quantum Hall regime (see Ref.~\onlinecite{nandi_nature_2012} and references therein). Finally, upturns of the Coulomb drag resistivity were reported in e-h doped GaAs/AlGaAs coupled quantum wells~\cite{croxall_prl_2008,seamons_prl_2009,morath_prb_2009}. However, the combination of 2d electron and hole gases in the same GaAs material required a large nanofabrication effort and the reported magnitude of the drag anomalies was smaller than that found in our hybrid heterostructures. Thus our observations establish a new class of vertical heterostructure devices with a potentially large flexibility in the design of band dispersions, doping, and e-h coupling where excitonic phenomena 
are easily accessible.

Systems of inter-layer excitons might be used to create coherent interconnections between electronic signal processing and optical communication in integrated circuits~\cite{high_science_2008,kuznetsova_opticsletters_2010,sanvitto_naturecommun_2013} or interfaced with superconducting contacts for a variety of applications, including analog-to-digital converters~\cite{dolcini_prl_2010} and topologically protected quantum bits~\cite{peotta_prb_2011}. The latter devices require, however, the exploitation of InAs-based 2DEGs that, unlike GaAs, make very good contact (i.e. no Schottky barriers) to superconductors~\cite{deon_apl_2011}. Inter-layer excitons in graphene/InAs hybrids may also pave the way for the exploration of the interplay between spin-orbit coupling and pairing fluctuations.

\acknowledgments 
We thank R. Duine, R. Fazio, A. Hamilton, M. Katsnelson, A. MacDonald, D. Neilson, K. Novoselov, A. Perali, A. Pinczuk, and G. Vignale for very useful discussions.
We acknowledge funding from EU Graphene Flagship (contract no. CNECT-ICT-604391), EC ITN project ``INDEX'' Grant No. FP7-2011-289968, the Italian Ministry of Education, University, and Research (MIUR) through the program ``FIRB - Futuro in Ricerca 2010'' Grant No.~RBFR10M5BT (``PLASMOGRAPH''), ERC grants NANOPOTS, Hetero2D, a Royal Society Wolfson Research Merit Award, EU projects GENIUS, CARERAMM, RODIN, EPSRC grants EP/K01711X/1, and EP/K017144/1.

\appendix

\section{Sample details and fabrication}
\label{appendix:sample}

\begin{figure}[t]
\center{
\includegraphics[width=\columnwidth]{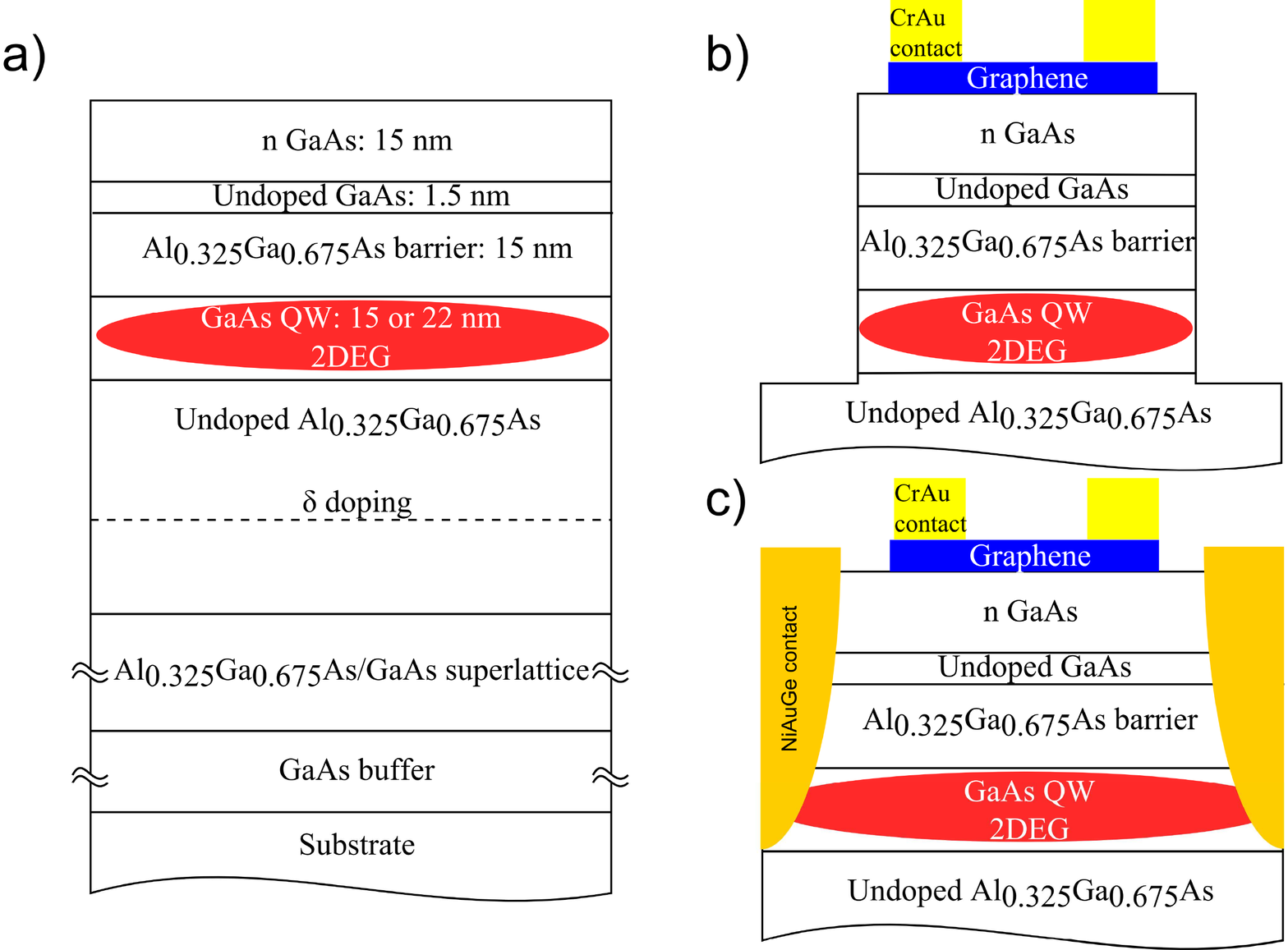}
\caption{a) Schematic of the GaAs/AlGaAs heterostructures (thicknesses of the layers are not to scale); b) section across the etched channel; c) section of the device along the channel.\label{fig:Sone}}
}
\end{figure}

The samples are modulation-doped GaAs/AlGaAs heterostructures hosting a 2DEG with single-layer (SLG) and bilayer graphene (BLG) flakes transferred onto them. Two different heterostructures (A and B) are investigated, differing only in the width of the quantum well, which lies $31.5~{\rm nm}$ from the surface in both cases. The layer sequence, Fig.~\ref{fig:Sone}a), starting from the surface comprises a $15~{\rm nm}$-thick $n$-doped GaAs cap layer, followed by a $1.5~{\rm nm}$ undoped GaAs and a $15~{\rm nm}$-thick undoped barrier layer of ${\rm Al}_{0.325}{\rm Ga}_{0.675}{\rm As}$. The GaAs quantum well has a thickness of $15~{\rm nm}$ in sample A and $22~{\rm nm}$ in sample B; it is followed by a thick ${\rm Al}_{0.325}{\rm Ga}_{0.675}{\rm As}$ barrier, which hosts a Si $\delta$-doping located $48.5~{\rm nm}$ from the well. The two samples differ in carrier density and mobility, as reported in the main text.
The Hall bar devices ($300~{\mu}{\rm m}$ wide and $1.500~{\mu}{\rm m}$ long) are fabricated by UV lithography. Ni/AuGe/Ni/Au metals are evaporated and annealed at $400^\circ~{\rm C}$ to form Ohmic contacts to the 2DEG. The mesa is then defined by wet etching in acid solution.

\begin{figure}[t]
\center{
\includegraphics[width=0.7\columnwidth]{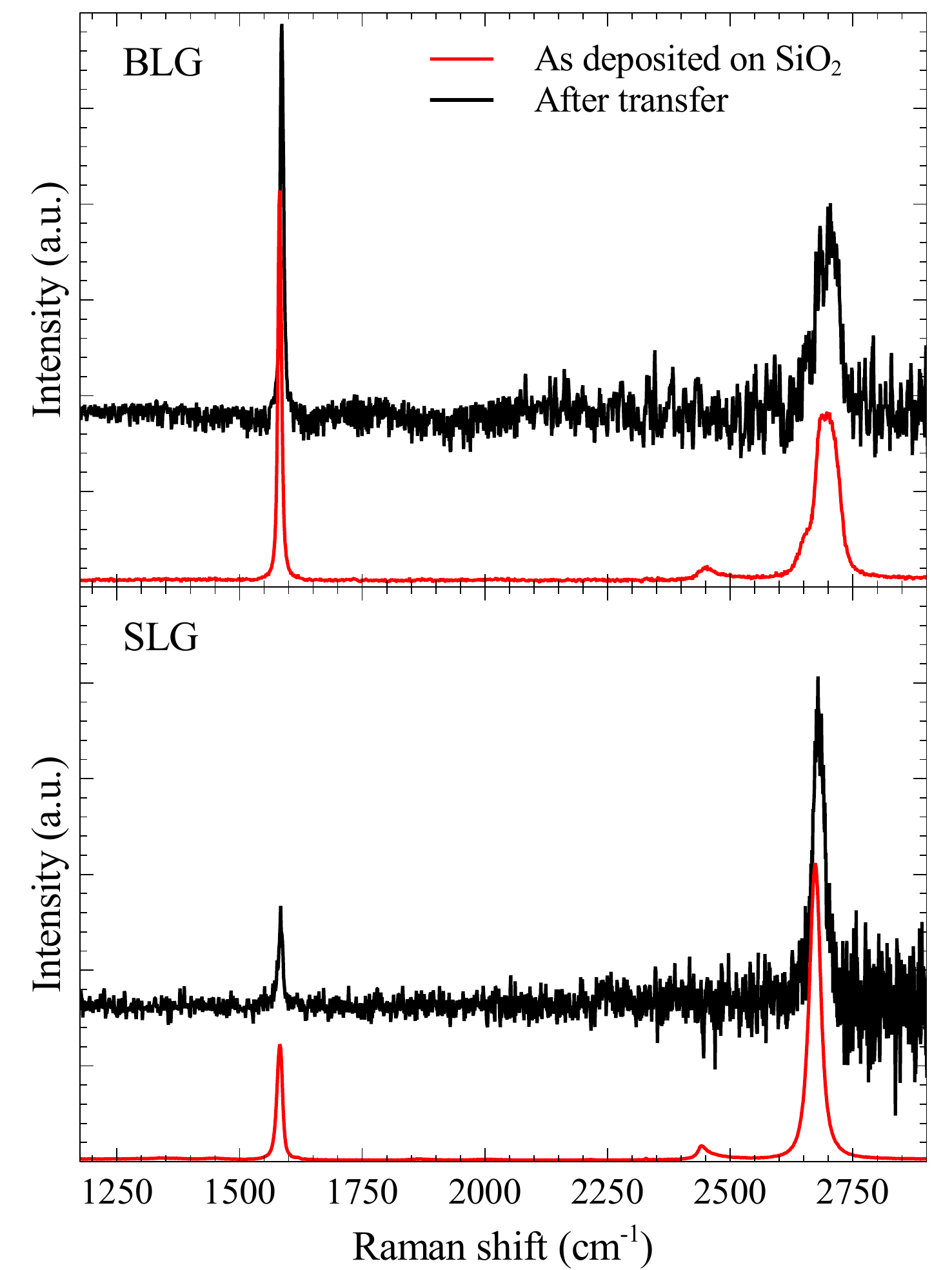}
\caption{Room-temperature Raman spectra of bilayer (BLG) (top panel) and single-layer (SLG) (bottom panel) graphene before and after transfer on the GaAs heterostructure.\label{fig:Stwo}}
}
\end{figure}

SLG and BLG flakes are produced by micromechanical exfoliation \cite{kostya_pnas_2005} of graphite on $\sim 300~{\rm nm}$ ${\rm SiO}_2$ on Si substrates. The number of layers is identified by a combination of optical microscopy~\cite{cnano} and Raman spectroscopy~\cite{RamanACF,RamanACF2}. The latter is also used to monitor the sample quality by measuring the D to G ratio~\cite{cancado} and the doping level~\cite{das}. Selected flakes are then placed onto a GaAs-based substrate at the center of a pre-patterned Hall bar by using a polymer-based wet transfer process~\cite{bonaccorso_matertoday_2012}. PMMA (molecular weight 950K) is first spin coated onto the substrate with micromechanically cleaved flakes, then the sample is immersed in de-ionized water, resulting in the detachment of the polymer film because of water intercalation at the PMMA-SiO$_2$ interface~\cite{bonaccorso_matertoday_2012}. Graphene flakes attach to the polymer and are removed from the ${\rm Si}/{\rm SiO}_2$ substrate. The polymer+graphene film is then placed onto the target substrate and, after complete drying of the water, PMMA is removed by acetone. Success of the transfer is confirmed by optical inspection (bright and dark field microscopy),  atomic force microscopy (AFM) and Raman spectroscopy. Raman spectra are collected with a Renishaw InVia spectrometer using laser excitation wavelengths at $514.5~{\rm nm}$. Excitation power is kept below $1~{\rm mW}$ to avoid local heating and the scattered light is collected with a 100X objective. Fig.~\ref{fig:Stwo} plots the spectra of SLG and BLG flakes before and after transfer. 
The strong luminescence background due to the GaAs/AlGaAs substrate has been subtracted out in order to highlight the Raman signal of the SLG/BLG flakes after transfer. This, combined with the lack of interference enhancement on the GaAs/AlGaAs substrate, explains why the spectra of the transferred flakes are noisy. As discussed in the main text for the SLG flake, also for the BLG flake we do not see any increase of D peak, thus showing that the transfer procedure does not induce extra defects.

Since the flakes are much smaller than the Hall bar widths, in each device we define a narrow channel ($\sim30~{\mu}{\rm m}$) in the Hall bar by electron beam lithography (EBL) and wet etching, see Figs.~\ref{fig:Sone}b)-c). To avoid exposure of the flakes to the electron beam, we took an optical image to align the EBL. This procedure is possible because SLG and BLG on this substrate become optically visible once coated by PMMA, see Fig.~\ref{fig:one}. Finally, Ohmic contacts (Cr/Au) are fabricated by EBL, metal evaporation and lift-off.

\section{Electrical characterization}
\label{appendix:characterization}
\begin{figure}[t]
\center{
\includegraphics[width=0.7\columnwidth]{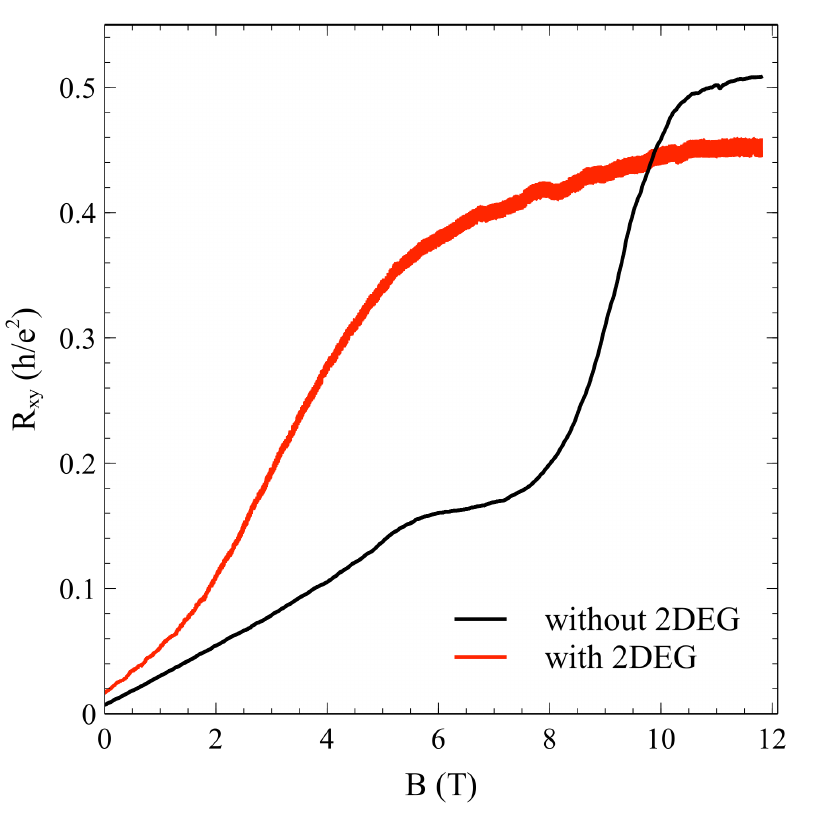}
\caption{Hall resistance as a function of perpendicular magnetic field in SLG at $T=4~{\rm K}$. The red curve corresponds to the case in which the 2DEG is induced by light illumination. The black curve corresponds to the case in which the GaAs quantum well is empty.\label{fig:Sthree}}
}
\end{figure}

As described in the main text, the anomalous quantum Hall effect (QHE) in SLG is seen at $4~{\rm K}$ when the 2DEG is {\it not} induced in the GaAs channel (see also black curve in Fig.~\ref{fig:Sthree}). A markedly different result is seen when the 2DEG is induced by LED illumination. This is shown in Fig.~\ref{fig:Sthree}, where we compare the Hall resistance at $T=4~{\rm K}$ as a function of the perpendicular magnetic field up to $12~{\rm T}$ for the two cases.
Because of the massless Dirac fermion nature of the charge carriers in SLG and the spin-valley degeneracy, plateaus are expected~\cite{castroneto_rmp_2009} in the Hall resistance at $h/(\nu e^2)$, with $\nu=2,6,10,\dots$. When the 2DEG is not induced, QHE plateaus are visible at $\nu = 2, 6$ while, as expected, the $\nu=4$ plateau is missing. In the other case, the plateau at $\nu = 6$ is not visible and the resistance approaches the value for $\nu =2$ at the highest magnetic field. The two configurations have different mobility, $4100~{\rm cm}^2/({\rm V}{\rm s})$, without the 2DEG, $2100~{\rm cm}^2/({\rm V}{\rm s})$ with the 2DEG. As explained in the main text, we can explain this difference considering that charged impurities (ionized Si donors) are left in the heterostructure when the 2DEG is induced, resulting in turn in an enhanced scattering of SLG carriers, which reduces mobility~\cite{dassarma_rmp_2011}.

\section{Inter-layer (``leakage'') current}
\label{appendix:interlayertransport}
Measurements of the leakage current between the two layers are performed by applying a voltage source to the 2DEG and detecting the current with an ammeter connected to SLG/BLG~\cite{Gupta}. The applied voltage is negative to avoid depletion of the 2DEG.
\begin{figure}[t]
\center{
\includegraphics[width=0.7\columnwidth]{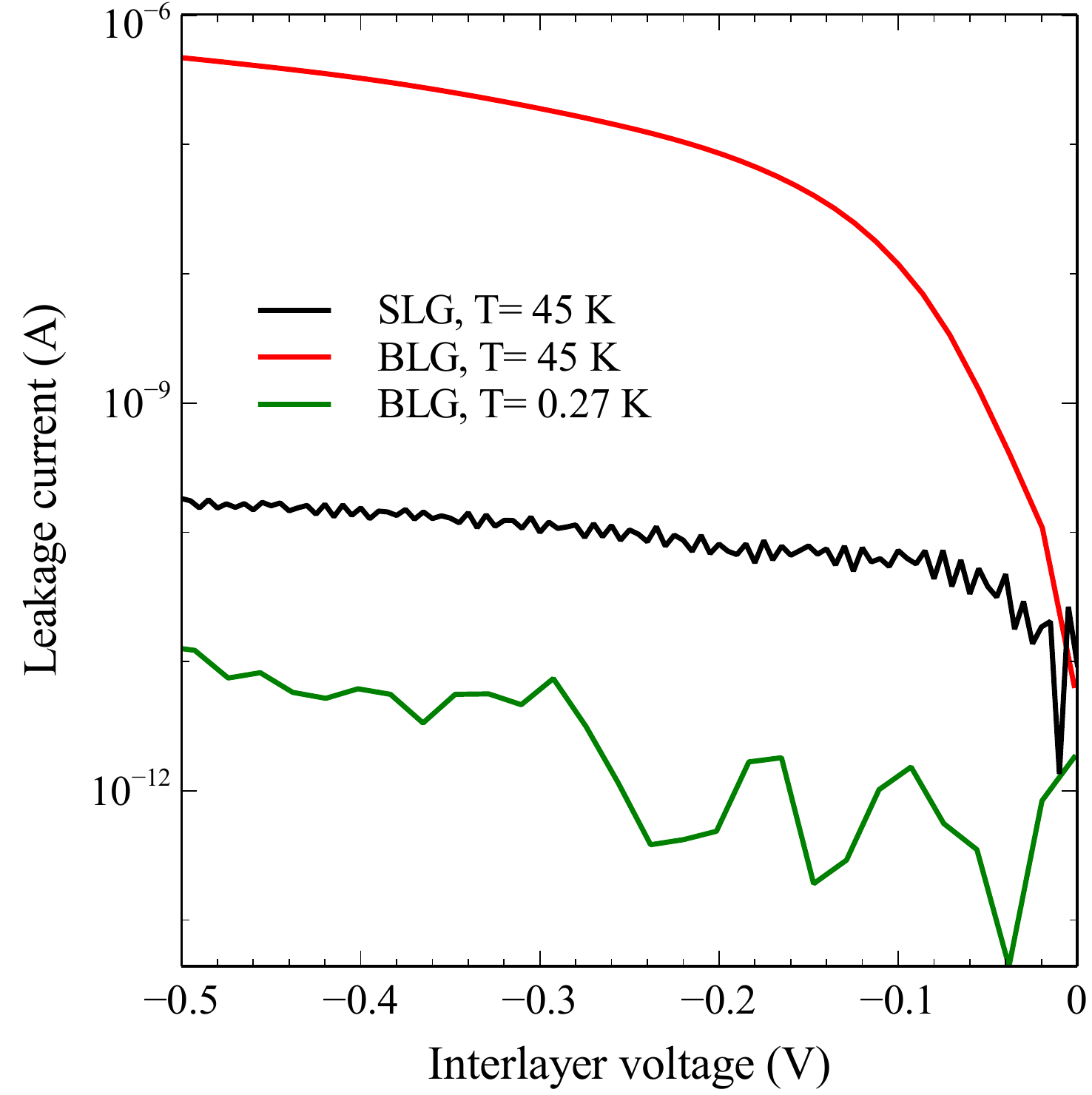}
\caption{Representative data for the leakage current between 2DEG and SLG/BLG as a function of the inter-layer voltage at different temperatures.\label{fig:Sfour}}
}
\end{figure}
We report in Fig.~\ref{fig:Sfour} the measured leakage currents for the SLG and BLG devices as a function of the inter-layer voltage. We recall that in the drag experiment configuration, the maximum negative value for this voltage is $\sim-0.3~{\rm V}$.  In the case of SLG, the current is smaller than $1~{\rm nA}$ even at $45~{\rm K}$. In the case of BLG we find a larger current at $45~{\rm K}$, but still much smaller than the drive current measured in the drag experiment (in the worst case of $-0.3~{\rm V}$ we have a leakage current of $\approx 100~{\rm nA}$ while the drive current is $2~{\mu}{\rm A}$). At the lowest temperature, the leakage current in BLG is smaller than the drive current by many orders of magnitude, so we do not expect it to affect the drag measurement~\cite{gorbachev_naturephys_2012}.

\section{Drag measured by using graphene as a passive layer}
\label{appendix:fluctuations}
\begin{figure}[t]
\center{
\includegraphics[width=\columnwidth]{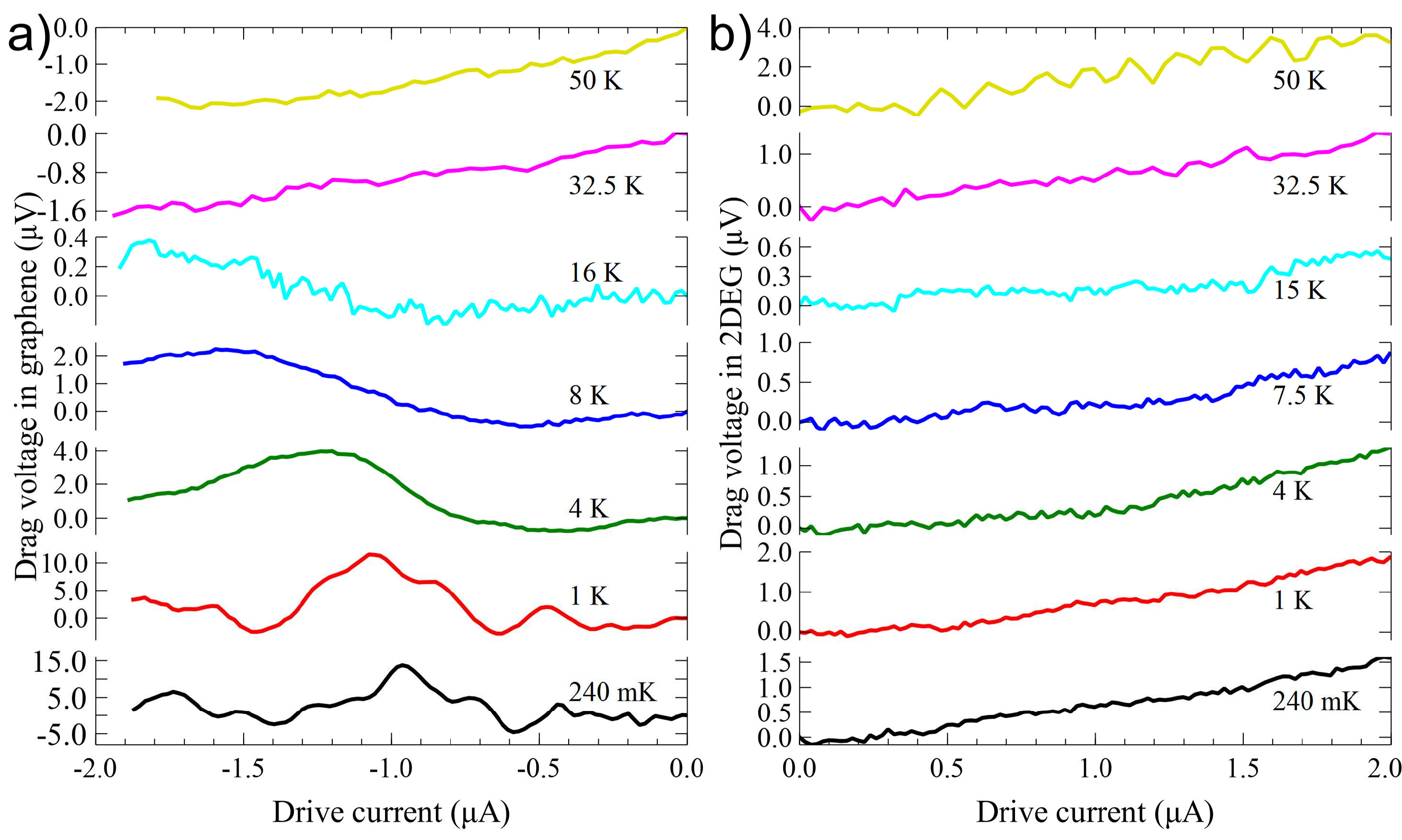}
\caption{Temperature dependence of the $V_{\rm drag}$-$I_{\rm drive}$ characteristics. a) The induced drag voltage measured in SLG is plotted as a function of the drive current injected in the 2DEG. b) The induced drag voltage measured in the 2DEG is plotted as a function of the drive current injected in SLG. Different traces refer to different values of the temperature, which decreases from top to bottom.\label{fig:Sfive}}
}
\end{figure}

The temperature dependence of $V_{\rm drag}$ in SLG as a function of the drive current $I_{\rm drive}$ in the 2DEG is reported in Fig.~\ref{fig:Sfive}a). Upon reducing temperature, the drag voltage measured in SLG displays a series of oscillations, which we believe to be linked to mesoscopic fluctuations already discussed in Coulomb drag setups based on two spatially-separated SLG sheets~\cite{gorbachev_naturephys_2012,kim_prb_2011,kim_ssc_2012} and also for all GaAs/AlGaAs double layers~\cite{price_science_2007}. These fluctuations disappear for $T\gtrsim 16~{\rm K}$. Above this temperature the $V_{\rm drag}$ - $I_{\rm drive}$ relation becomes linear.
When the voltage drop is measured in the 2DEG no fluctuations arise at all the explored temperatures: see Fig.~\ref{fig:Sfive}b). This allows us to extract the evolution of the drag resistance down to $250~{\rm mK}$ (see main text). A similar fluctuating behaviour of the drag voltage at low temperature is found in sample B in the configuration in which BLG is used as the passive layer.

\section{Theoretical background}
\label{appendix:theory}

We summarize the most elementary theory of drag resistance in the Fermi-liquid regime~\cite{rojo_jpcm_1999}, which is based on Boltzmann-transport theory supplemented by second-order perturbation theory in the screened inter-layer interaction~\cite{rojo_jpcm_1999}. 

In the SLG/2DEG vertical heterostructure, the drag resistivity $\rho_{\rm D}$ is given, in the low-temperature limit, by~\cite{principi_prb_2012}:
\begin{eqnarray}\label{eq:rho_hybridT0}
\rho_{\rm D}(T) &=& - \frac{1}{24 e^2} \frac{1}{v_{{\rm F,t}} v_{{\rm F, b}}}\frac{(k_{{\rm B}}T)^2}{ \varepsilon_{{\rm F, t}}\varepsilon_{{\rm F, b}}}  \int_0 ^{q_{{\rm max}}} q ~ d q | U_{\rm tb}(q,0)|^2 \nonumber\\
&\times&{\cal F}\left(\frac{q}{2 k_{\rm F, t}}, \frac{q}{2 k_{\rm F, b}}\right)~,
\end{eqnarray}
where $v_{\rm F,t}$ ($v_{\rm F,b}$) is the Fermi velocity in the top (bottom) layer, $\varepsilon_{\rm F, t}$ ($\varepsilon_{\rm F, b}$) is the Fermi energy in the top (bottom) layer, $k_{\rm F,t}$ ($k_{\rm F,b}$) is the Fermi wave number in the top (bottom) layer, $q_{{\rm max}} = \min( 2 k_{{\rm F, t}} , 2k_{{\rm F, b}})$, and ${\cal F}(x,y) = \sqrt{(1-x^2)/(1-y^2)}$.  The low-temperature limit is defined by the inequality $k_{{\rm B}} T \ll \min (\varepsilon_{{\rm F},{\rm t}}, \varepsilon_{{\rm F},{\rm b}})$.

In Eq.~(\ref{eq:rho_hybridT0}) the screened interaction $U_{\rm tb}$ is given by:
\begin{equation}\label{eq:screenedinteraction}
U_{\rm tb}(q,\omega) = \frac{V_{\rm tb}(q)}{\varepsilon (q,\omega)}~,
\end{equation}
where
\begin{eqnarray}\label{eq:zeroes}
\varepsilon(q,\omega) &=& [1-V_{\rm tt}(q) \chi_{\rm t}(q,\omega)][1-V_{\rm bb}(q)\chi_{\rm b}(q,\omega)] \nonumber\\
&-& V^2_{\rm tb}(q)\chi_{\rm t}(q,\omega)\chi_{\rm b}(q,\omega)
\end{eqnarray}
is the dielectric function in the random phase approximation~\cite{Giuliani_and_Vignale}. In Eq.~(\ref{eq:zeroes}), $\chi_{\rm t}(q,\omega)$ and $\chi_{\rm b}(q,\omega)$ are the density-density linear response functions of the electronic fluids in the top and bottom layer, respectively. Microscopic expressions for the density-density response function $\chi_{\rm t}(q,\omega)$ of the electron fluid in a doped graphene sheet can be found in Refs.~\onlinecite{wunsch_njp_2006,hwang_prb_2007,barlas_prl_2007}. The density-density response function $\chi_{\rm b}(q,\omega)$ of a 2DEG is extensively discussed in Ref.~\onlinecite{Giuliani_and_Vignale}.

In Eqs.~(\ref{eq:screenedinteraction})-(\ref{eq:zeroes}), $V_{\rm tt}(q)$ is the Coulomb interaction between two charges in the top layer,
\begin{equation}\label{eq:vtt}
V_{\rm tt}(q) = \frac{4\pi e^2g(q)}{q (\epsilon_1+\epsilon_2)}~,
\end{equation}
while $V_{\rm bb}(q)$ is the Coulomb interaction in the bottom layer,
\begin{equation}\label{eq:vbb}
V_{\rm bb}(q) = \frac{4\pi e^2 g(q)}{q D(q)} [ (\epsilon_2 + \epsilon_1) e^{qd} +
 (\epsilon_2 - \epsilon_1) e^{-qd}]~,
\end{equation}
with $D(q) =  2\epsilon_2(\epsilon_1 + \epsilon_2) e^{qd}$. Finally, the inter-layer interaction is given by
\begin{equation}\label{eq:vtb}
V_{\rm tb}(q) = V_{\rm bt}(q) = \frac{8\pi e^2}{q D(q)}~\epsilon_2g(q)~.
\end{equation}
The dimensionless parameter $\epsilon_1$ represents the relative dielectric constant of the material above SLG (in our case air, $\epsilon_1 = 1$), while $\epsilon_2$ is the relative dielectric constant of GaAs, which is $\sim 13$. In writing Eqs.~(\ref{eq:vtt})-(\ref{eq:vtb}) we neglect the difference between the dielectric constant of GaAs and AlGaAs. Note that in Eqs.~(\ref{eq:vtt})-(\ref{eq:vtb}) we introduced a form factor $g(q)<1$, which stems from the finite width of the quantum well hosting the 2DEG. This can be found~\cite{davies} by solving the Poisson equation for the SLG/2DEG vertical heterostructure under the assumption that the confining potential for the 2DEG along the growth direction is given by a square quantum well of width $w$. This assumption has been checked with the help of a self-consistent Poisson-Schr\"{o}dinger solver. We find
\begin{equation}
g(q) = \frac{1 - \exp{( - w q)}}{w q + w^3 q^3 /(4 \pi^2)}~.
\end{equation}


\begin{thebibliography}{77}
\bibitem{footnote_temperature} E.g., fixing $T_{\rm c} = 10~{\rm mK}$, the best fit yields $A = (0.416 \pm 0.015)~\Omega$ and $R_0 = (2.66 \pm 0.08)~\Omega$. We can well fit the experimental data by choosing any value of $T_{\rm c}$, as long as this is substantially lower than the lowest investigated temperature, i.e.~$T= 240~{\rm mK}$. As it is clear from the functional form of the fitting function in Eq.~(\ref{eq:logfit}), in this regime a change $T_{\rm c} \to T'_{\rm c}$ can be reabsorbed into a change of the background resistance $R_0 \to R_0 + A\log(T'_{\rm c}/T_{\rm c})$.
\end{thebibliography}
\end{document}